\documentstyle[epsf,prc,preprint,aps]{revtex}
\def\be{\begin{equation}}
\def\ee{\end{equation}}
\def\bea{\begin{eqnarray}}
\def\eea{\end{eqnarray}}
\def\bP{\bar{\Psi}}
\def\pm{\partial_\mu}
\def\gm{\gamma_\mu}
\def\ggm{\gamma_\mu \gamma_5}
\def\vp{\vec {\Pi}}
\def\vv{\vec {\cal V}_\nu}
\def\va{\vec {\cal A}_\mu}
\def\vt{\vec {\tau}}
\begin{document}
%Omit the next line when submitting
\tighten
\draft
\title
{\bf Comments on the last development in constructing the amplitude for the
radiative muon capture
}
\author{J.\,Smejkal and E.\,Truhl\'{\i}k}
\address{
Institute of Nuclear Physics, Czech Academy of Sciences, CZ-25068
$\check{R}$e$\check{z}$, Czech Republic\\
%\vspace{-15pt}
%\begin{center}
%and
%\end{center}
%\vspace{-30pt}
}
%\author{F.\,C.\,Khanna
%}
%\address{Theoretical Physics Institute, Department of Physics, University
%of Alberta, Edmonton, Alberta,Canada,T6G 2J1\\
%\vspace{-15pt}
%\begin{center}
%and
%\end{center}
%\vspace{-15pt}
%TRIUMF, 4004 Wesbrook Mall, Vancouver, BC, Canada, V6T 2A3\\}
\maketitle
\begin{abstract}
%ABSTRACT%%%%%%%%%%%%%%%%%%%%%%%%%%%%%%%%%%%%%%%%%%%%%%%%%%%%%%%%%
It has recently been claimed by Cheon and Cheoun that the discrepancy between the 
experimental value of the induced pseudoscalar $g_P$, obtained recently at 
TRIUMF from a measurement of the radiative muon capture by proton, and its  
value predicted by using PCAC and pion pole dominance can be explained
by a contact term generated from a simple pion-nucleon Lagrangian. We show
in our comment that this claim is ill founded. 
\end{abstract}
\newpage
In a recent preprint \cite{CC}, Cheon and Cheoun claim to
remove the discrepancy between the value of the induced pseudoscalar $g_P$ 
obtained in the TRIUMF experiment \cite{TRIUMF} investigating the radiative muon
capture (RMC) by proton and its prediction from the PCAC and the pion pole dominance.
They start from a linear $\sigma$-model Lagrangian of the $\pi-N$ system
and derive a general equation for the axial current and its divergence. 
Then they introduce into the 
Lagrangian the electromagnetic coupling by the minimal substitution 
$\partial_\mu\,\longrightarrow\,D_\mu\,=\,\partial_\mu\,-\,ie\epsilon_\mu$ and using 
a standard local transformation of the nucleon $\Psi(x)$ and pion 
$\phi_a(x)$ fields,
\be
\Psi\,\rightarrow\,\Psi'\,=\,(1+i\gamma_5 \frac{\vec{\tau}}{2} \cdot \vec{\eta}(x))
\Psi\,,\qquad\,\vec{\phi}\,\rightarrow\,\vec{\phi}'\,=\,\vec{\phi}-f_\pi 
\vec{\eta(x)}\,,   \label{tran}
\ee
and the above mentioned equation of motion, they derive the axial current 
$A^\mu_a(x)$ in the presence
of the electromagnetic field. In the final Eq.\,(22) for the axial current, 
besides the well known terms a 
large contact term (the last term in the r.\,h.\,s.\, of the equation) appears
which, according to Cheon and Cheoun, should resolve the problem.

The basic equation of Ref.\,\cite{CC} is Eq.\,(14)
\be
{\widetilde {\cal L}}_0\,=\,\bar{\Psi}[i\gamma^\mu\,D_\mu+gf_\pi exp(\frac{i}{f_\pi}
\gamma_5 \vec{\tau} \cdot \vec{\phi})]\Psi\,-\,\bar{\Psi}\gamma^\mu\gamma_5
\frac{\vec{\tau}}{2}\Psi \cdot (D_\mu\vec{\eta})\,+\,\frac{1}{2}(D_\mu\vec{\phi})^2
-(D^\mu\vec{\phi}) \cdot f_\pi (D_\mu \vec{\eta})\,.
\ee
Actually, it is the 3rd term in the r.\,h.\,s.\, of this equation
$\sim\,(D_\mu \vec{\eta})$ which leads to the large contact contribution. But it can be 
seen that under the local transformations (\ref{tran}), the correct term is
\be
-\Psi \gamma^\mu \gamma_5 \frac{\vec{\tau}}{2}\Psi \cdot (\partial_\mu \vec{\eta})
\,=\,-\Psi \gamma^\mu \gamma_5 \frac{\vec{\tau}}{2}\Psi \cdot (D_\mu \vec{\eta})
\,-\,\Psi \gamma^\mu \gamma_5 \frac{\vec{\tau}}{2}\Psi \cdot (ie \epsilon_\mu 
\vec{\eta})\,.   \label{CT}
\ee
Using Eqs.\,(13) \cite{CC} defining the current and its divergence and Eq.\,(14) 
\cite{CC} with the new 
term  $\sim\,(ie \epsilon_\mu \vec{\eta})$ from our Eq.\,(\ref{CT}) included 
we can see that instead of the current $A^\mu_a(x)$ with the divergence equal 
zero we now have [cf.\, Eqs.\,(16) \cite{CC}]
\be
D^{(+)}_\mu\,A^\mu_a\,=\,\Psi \gamma^\mu \gamma_5 \frac{\tau_a}{2}\Psi 
(ie\epsilon_\mu)\,.   \label{DIV}
\ee
Taking into account  Eq.\,(\ref{DIV}) and following the derivation of the axial current
in \cite{CC} we get the same Eq.\,(22) but with the last term in the r.\,h.\,s.\,
multiplied by the factor \mbox{$(g_A\,-\,1)/g_A\,\approx\,0.20$}
\be
        \frac{g_A-1}{g_A}\frac{eg_P(q^2)}{2mm_\mu}\,q^\mu\,
        [\bP(x)\,\epsilon_\alpha \gamma^\alpha \gamma_5\frac{\tau_a}{2}\,\Psi(x)]\,.
\label{CTS}
\ee
 So the effect of the
questioned contact term is strongly suppressed.

Actually, the appearance of such a term in the leading order in the RMC amplitude 
constructed at the tree level  from a chiral invariant Lagrangian 
would contradict the low energy theorem prediction for the hadron part of
the RMC amplitude and in essence the current algebras and PCAC. 
It appears in \cite{CC} due to an artificial introduction of a piece of the pseudovector 
 $\pi-N$ coupling via Eq.\,(18) which provides the factor
$(g_A\,-\,1)/g_A$. Otherwise, the current (\ref{CTS}) should be absent in 
any model based on the chiral
Lagrangian with any mixing of the $\pi-N$ couplings (see below).  

We have recently  derived \cite{STK} the RMC amplitude from a chiral invariant 
Lagrangian of the $N \pi \rho\,a_1\,\omega$ system constructed within the framework
of the hidden local symmetry approach \cite{BKY,M,KM,STG}.
The contact term of the type just discussed above appears in our Eq.\,(3.11) but
it is cancelled in the leading order by another contact term of Eq.\,(3.32) and
only higher order terms in k and q survive. The chiral Lagrangian of the
$\pi-N$ system can be obtained from our Lagrangian Eq.\,(2.1) by 
performing the limit $m_B\,=\,\infty$ for all heavy mesons. As an example, we
give here the $N \pi$ Lagrangian with the pseudovector $\pi-N$ coupling
\bea
{\cal L}_{N\pi}\,&=&\,-\bP\,\gm\pm\,\Psi\,-\,M\,\bP\,\Psi\,-\,i\frac{1}{4f^2_\pi}\,
\bP\,\gm \vt\,\Psi \cdot (\vp \times \pm \vp)  \nonumber \\
& &\,-i\frac{e}{2}\,\bP\,\gm \vec {\cal V}_\mu \cdot \vt\,\Psi\,-\,i\frac{e}{2f_\pi}\,
\bP\,\gm \vt\,\Psi \cdot (\vp \times \va)  \nonumber  \\
& & \,-\frac{e}{4}\frac{\kappa_V}{2M}\,\bP\,\sigma_{\mu \nu} \vt\,\Psi \cdot
(\pm \vv\,-\,\partial_\nu \vec {\cal V}_\mu)\,
+\,\frac{e^2}{4}\frac{\kappa_V}{2M}\,\bP\,\sigma_{\mu \nu} 
\vt\,\Psi \cdot(\vec {\cal V}_\mu \times \vv)  \nonumber  \\
& & \,-i\frac{g_A}{2f_\pi}\,\bP\,\ggm \vt\,\Psi \cdot (\pm \vp)\,-\,
i\frac{eg_A}{2}\,\bP\,\ggm \vt\,\Psi \cdot \va  \nonumber  \\
& & \,-i\frac{e g_A}{2f_\pi}\,\bP\,\ggm \vt\,\Psi \cdot ( \vp \times 
\vec {\cal V}_\mu)\,+\,
{\cal O}(|\Psi|^4\,,|\Pi|^3)\,,   \label{LNpi}
\eea
\bea
{\cal L}_\pi\,&=&\,-\frac{1}{2}(\pm \vp)^2\,-\,ef_\pi(\va \cdot \pm \vp)\,+\,
e\,\vec {\cal V}_\mu \cdot (\vp \times \pm \vp) \nonumber \\
& & \,-e^2\,f_\pi \vp \cdot (\vec {\cal V}_\mu \times \va)\,+\,
{\cal O}(|\Pi|^4)\,.  \label{Lpi}
\eea
The associated currents derived by the Glashow--Gell-Mann method \cite{GGM} read
\bea
\vec {J}_{V,\,\mu}\,&=&\,-(\vp \times \pm \vp)\,-\,e f_\pi\,(\vp \times \va)
\,+\,\frac{i}{2}\,\bP\,\gm \vt\,\Psi 
 \,-\,i\frac{g_A}{2f_\pi}\,\bP\,\ggm (\vp \times \vt)\,\Psi \nonumber  \\
& & -\,\frac{\kappa_V}{4M}\,\partial_\nu[\bP\,\sigma_{\mu \nu} \vt\,\Psi]
\,-\,\frac{e}{2}\frac{\kappa_V}{2M}\,\bP\,\sigma_{\mu \nu} 
(\vv \times \vt)\,\Psi  
 \,+\,{\cal O}(|\Psi|^4\,,|\Pi|^3)\,,   \label{JV}
\eea
\bea
\vec{J}_{A,\,\mu}\,&=&\,f_\pi\, \pm \vp\,+\,e f_\pi\,(\vp \times \vec {\cal V}_\mu)\,+\,i
\frac{g_A}{2}\,\bP\,\ggm \vt\,\Psi  \nonumber \\
& & -\,i\frac{1}{2f_\pi}\,\bP\,\gm(\vp \times \vt)\,\Psi\,+\,
{\cal O}(|\Psi|^4\,,|\Pi|^3)\,.  \label{JA}
\eea
Here $\vv$ and $\va$ are the external vector and axial fields.

The only seagull of the type (\ref{CTS}) can be constructed using the last
term of the Lagrangian (\ref{LNpi}) and the first term of the current
(\ref{JA}), but it does not contribute to the RMC
amplitude. The same situation can be seen after applying the Foldy-Dyson
transformation to the Lagrangian (\ref{LNpi})
\be
\Psi\,=\,exp(-i\lambda \frac{g_A}{2f_\pi}\gamma_5\,(\vt \cdot \vp))\,\Psi'\,,
\ee
which would yield the Lagrangian with the  $\pi-N$ couplings mixed (for
$\lambda\,=\,1$, one gets the pseudoscalar one).

It has recently been argued by Fearing \cite{F} that the term (\ref{CTS}) violates
the gauge invariance of the RMC amplitude (26) derived in \cite{CC}, which is somewhat 
misleading. 
Actually, its presence would violate the CVC constraint which should satisfy 
the hadron part of this amplitude. The validity of this
constraint then guarantees the gauge invariance of the whole RMC amplitude 
\cite{ChS,STK}.
On the contrary, the  presence of the term (\ref{CTS}) would violate directly the PCAC 
constraint for the hadron part of the RMC amplitude (26) \cite{CC}.
%\narrowtext
%\input feynman
%\input{intro}
%\input{chapter1}
%\input{chapter2}
%\input{chapter3}
%\acknowledgments
%{\bf Acknowledgments}
\vspace{10pt}

This work is supported  by the grant GA \v{C}R 202/97/0447.

%LITERATURE%%%%%%%%%%%%%%%%%%%%%%%%%%%%%%%%%%%%%%%%%%%%%%%%%%%

%\input{fig}
\end{document}